\documentclass[12pt]{article}

\begin{document}
\begin{center}
{\bf Quantization of bosonic fields with two mass and spin states}

\vspace{5mm}
 S. I. Kruglov\\

\vspace{5mm}
\textit{University of Toronto at Scarborough,\\ Physical and Environmental Sciences Department, \\
1265 Military Trail, Toronto, Ontario, Canada M1C 1A4}
\end{center}

\begin{abstract}
We investigate bosonic fields possessing two mass and spin states.
The density matrix in the first order
formalism is obtained. The quantization of fields in the
first order formulation is performed and propagators are found.
\end{abstract}

\vspace{5mm}
Pacs: 03.65.Pm, 03.70.+k, 04.20.Fy, 11.10.-z

\section{Introduction}

The theory of massive vector fields play very important role in the standard
electroweak model. We mention also that propagating torsion axial vector fields
appear in quantum gravity \cite{Buchbinder}. These fields possess
two mass and spin states \cite{Kruglov}. The classical theory of
bosonic fields with two mass and spin states was considered in \cite{Kruglov2}.
Here we quantize such fields in the first order formalism.

The paper is organized as follows. In Sec.2 the propagator of the
bosonic fields with two mass and spin states in the second order
formulation is obtained. The first order wave equation for fields
in the 11$\times$11-matrix form is investigated in Sec.3. We obtain
the minimal equation for the 11$\times$11-matrices and the density matrix.
The quantization of the field is discussed in Sec.4. The commutation
relations and the vacuum expectation of chronological pairing of
operators are found. A conclusion is given in Sec.5.

We use the Euclidian metric and the four-vector of the field is
$B_\mu=(B_m,B_4)$; $m=1,2,3$; $B_4=iB_0$.

\section{Field equations}

It the work \cite{Kruglov2}, we considered charged bosonic fields which
possess two spins, one and zero, with different masses.
The corresponding field equations are
\begin{equation}
\partial_\mu^2 B_\nu +\left( \frac{m^2}{m_0^2}-1\right) \partial _\nu \left(
\partial_\alpha B_\alpha \right) -m^2 B_\nu =0,  \label{1}
\end{equation}
where $m$, $m_0$ are masses of spin-1 and spin-0 states, correspondingly.
For the real functions $(\textbf{B},B_0)$ Eq.(1) describes torsion fields
that appear in higher derivative quantum gravity \cite{Kruglov}.
Eq.(1)  can be represented, in momentum space, as 4$\times$4-matrix
equation:
\begin{equation}
M_{\mu\nu}B_\nu
=0,~~~~M_{\mu\nu}=\left(p^2+m^2\right)\delta_{\mu\nu}+
\left(\frac{m^2}{m_0^2}-1\right)p_\mu p_\nu . \label{2}
\end{equation}
To have non-trivial solutions of  Eq.(1), one gets det $M=0$. This
equation leads to the dispersion equation \cite{Kruglov2}:
\begin{equation}
\left( p^2+m^2\right)\left( p^2+m_0^2\right) =0,  \label{3}
\end{equation}
where the four-momentum being $p_\mu=(\textbf{p},ip_0)$.

The propagator of fields is defined as
$\left(M_{\mu\nu}\right)^{-1}$. We obtain
\begin{equation}
\left(M_{\mu\nu}\right)^{-1}=\frac{p^2+m_0^2+\left(m_0^2/m^2-1\right)p_\mu
p_\nu}{\left( p^2+m^2\right)\left( p^2+m_0^2\right)} ,  \label{4}
\end{equation}
so that
$M_{\mu\alpha}\left(M_{\alpha\nu}\right)^{-1}=\delta_{\mu\nu}$. The
propagator (4) corresponds to bosonic fields that have two states
with spin-1, mass $m$, and spin-0, mass $m_0$. It is convenient to
represent the propagator (4) as a sum of two propagators
\begin{equation}
\left(M_{\mu\nu}\right)^{-1}=\frac{1+\left(1/m^2\right)p_\mu
p_\nu}{p^2+m^2}- \frac{\left(1/m^2\right)p_\mu p_\nu}{p^2+m_0^2}.
\label{5}
\end{equation}
The first term in Eq.(5) is the propagator of the Proca equation
which can be obtained from Eq.(1) at $m_0\rightarrow \infty$. At
this limit ($m_0\rightarrow \infty$) the second term in Eq.(5)
vanishes and, therefore, corresponds to the contribution of spin-0
state. The second term in Eq.(5) has a "wrong" sign $(-)$ that
indicates on the presence of a ghost. The ghost results in the
negative contribution to the energy, and we should introduce
indefinite metrics in quantum field theory \cite{Kruglov3}.
Therefore, the state with spin-0 and mass $m_0$ may be considered as
a ghost and can be removed at the limit $m_0\rightarrow \infty$. We may treat,
therefore, Eq.(1) as the equation for the vector fields in the general gage \cite{Ruegg}
(see also references therein).

\section{The first order wave equation}

In \cite{Kruglov2}, Eq.(1) was represented
in 11-component matrix form as follows:
\begin{equation}
\left( \alpha _\mu \partial _\mu +m P_1+\frac{m^2_0}{m} P_0\right)
\Psi (x)=0, \label{6}
\end{equation}

\begin{equation}
\Psi (x)=\left\{ \psi _A(x)\right\} =\left(
\begin{array}{c}
(1/m)B(x) \\
B _\mu (x) \\
(1/m)B _{\mu \nu}(x)
\end{array}
\right),
\label{8}
\end{equation}
where $B_{\mu\nu}=\partial_\mu B_\nu-\partial_\nu B_\mu $,
$B (x)=-(m^2/m_0^2)\partial_\mu B_\mu(x)$.
Matrices entering Eq.(6), expressed trough the elements of the
entire matrix algebra $\varepsilon ^{A,B}$, are given by
\[
\alpha _\mu =\beta _\mu ^{(1)}+\beta _\mu ^{(0)},~~\beta _\mu
^{(1)}=\varepsilon ^{\nu ,[\nu \mu ]}+\varepsilon ^{[\nu \mu ],\nu
},~~\beta _\mu ^{(0)}=\varepsilon ^{\mu ,0}+\varepsilon ^{0,\mu },
\]
\vspace{-8mm}
\begin{equation}
\label{8}
\end{equation}
\vspace{-8mm}
\[
P_1=\varepsilon ^{\mu ,\mu }+\frac 12\varepsilon ^{[\mu \nu ],[\mu
\nu ]},\hspace{0.5in}P_0=\varepsilon ^{0,0}.
\]
It is convenient to consider the equivalent form of Eq.(6).
Multiplying Eq.(6) by the matrix
\[
M\left(m P_1+\frac{m^2_0}{m}
P_0\right)^{-1}=M\left(\frac{1}{m}P_1+\frac{m}{m_0^2}P_0\right),
\]
where
\begin{equation}
M=\frac{mm_0}{m+m_0}\label{9}
\end{equation}
is a reduced mass, and using the properties of the projection
operators $P_1^2=P_1$, $P_0^2=P_0$, we arrive at the standard form
of the first order relativistic wave equation
\begin{equation}
\left(\Gamma_\mu\partial_\mu+M\right)\Psi (x)=0. \label{10}
\end{equation}
The matrices of Eq.(10) are given by
\begin{equation}
\Gamma_\mu=\frac{m_0}{m+m_0}\left(\varepsilon ^{\mu ,0} +\varepsilon
^{\nu ,[\nu \mu ]}+\varepsilon ^{[\nu \mu ],\mu}\right)
+\frac{m^2}{m_0(m+m_0)}\varepsilon ^{0,\mu}, \label{11}
\end{equation}
and the wave function $\Psi (x)$ is the same as in Eq.(7). The
Hermitianizing matrix, obeying equations: $\eta \Gamma _i=-\Gamma^+
_i\eta^+$, $\eta \Gamma _4=\Gamma^+ _4\eta^+$ ($i=1,2,3$) is given
by
\begin{equation}
\eta =-\frac{m_0(m+m_0)}{m^2}\varepsilon
^{0,0}+\frac{m+m_0}{m_0}\left(\varepsilon ^{m,m}-\varepsilon
^{4,4}+\varepsilon ^{[m4],[m4]}-\frac{1}{2}\varepsilon ^{[m n ],[m
n]}\right). \label{12}
\end{equation}
The ``conjugated" wave function
$\overline{\Psi} (x)=\Psi^+(x)\eta$
obeys the equation as
follows:
\begin{equation}
\overline{\Psi} (x)\left( \Gamma _\mu \overleftarrow{\partial} _\mu
-M\right) =0. \label{13}
\end{equation}
In the momentum space Eq.(10) becomes
\begin{equation}
\left(i\hat{p}\pm  M \right)U_s(\pm p)=0 , \label{14}
\end{equation}
where $\hat{p}=\Gamma_\mu p_\mu$. Index $s$ in Eq.(14) corresponds
to the state with definite spin (one and zero) and spin projections
(for spin-1). One can verify that the matrix $\hat{p}$ satisfies the
minimal equation
\begin{equation}
\hat{p}^5 -\frac{\left(m^2+m_0^2\right)p^2}{\left(m+m_0\right)^2}
\hat{p}^3 + \frac{M^2 p^4 }{\left(m+m_0\right)^2} \hat{p}=0 .
\label{15}
\end{equation}
The squared of four-momentum $p^2$ is not specified here and obeys
the dispersion equation (3). Eq.(15) can be cast also in the form
\begin{equation}
\hat{p}\left(\hat{p}^2- \frac{M^2}{m_0^2}p^2\right) \left(\hat{p}^2
-\frac{M^2}{m^2}p^2\right)=0 . \label{16}
\end{equation}
Equations (15), (16) allow us to obtain the density matrix. With the
help of equations (15), (16), one can verify that the projection
matrix
\begin{equation}
\rho(\pm p)=\frac{\left(m+m_0\right)^4 i \hat{p}\left( i\hat{p}\mp
M\right)}{2m^2m_0^2\left(p^4-m^2m_0^2\right)} \left[ \widehat{p}^2+
\frac{p^4}{\left(m+m_0\right)^2}\right] \label{17}
\end{equation}
obeys the equations
\begin{equation}
\left(ip\pm  M \right)\rho(\pm p)=0 , \label{18}
\end{equation}
\begin{equation}
\rho(\pm p)^2=\rho(\pm p) ,~~~~\rho(+p)\rho(-p)=0. \label{19}
\end{equation}
The projection matrix (17) is the density matrix for impure spin
states:
\begin{equation}
\left(\rho(\pm p)\right)_{AB}=\sum_s \left(U_s(\pm p)\right)_A
\left(\overline{U}_s(\pm p)\right)_B,\label{20}
\end{equation}
where the $U_s(\pm p)$ is the wave function for the pure spin state
and obeys Eq.(14). It should be noted that the density matrix (17)
obtained obeys Eq.(18) for the arbitrary momentum $p_\mu$ satisfying
the dispersion equation (3). Contrarily, in
\cite{Kruglov2}, we found the density matrices only for the case
specifying the four-momentum: for the vector state $p^2=-m^2$ and
for the scalar state $p^2=-m_0^2$. Thus, we can introduce the
quantum number $\tau=0,1$, that $p^2=-m_\tau^2$ (no summation in
index $\tau$). It is implied that four momentum
$p=(\textbf{p},ip_0)$ possesses the additional quantum number
$\tau=0,1$ corresponding to the mass $m_0$ and $m\equiv m_1$.

From the Lagrangian
\begin{equation}
{\cal L}=-\frac{1}{2}\left[\overline{\Psi}(x)\left(\Gamma_\mu
\partial _\mu +M\right) \Psi (x)-\overline{\Psi}(x)\left(\Gamma_\mu
\overleftarrow{\partial} _\mu -M\right) \Psi (x)\right] ,\label{21}
\end{equation}
we obtain the momenta
\begin{equation}
\pi (x)=\frac{\partial\mathcal{L}}{\partial(\partial_0)\Psi
(x)}=\frac{i}{2}\overline{\Psi}(x)\Gamma_4,
\label{22}
\end{equation}
\begin{equation}
\overline{\pi}(x)=\frac{\partial\mathcal{L}}{\partial(\partial_0)\overline{\Psi}
(x)}=-\frac{i}{2}\overline{\Psi}(x)\Gamma_4
 .\label{23}
\end{equation}
The Hamiltonian (energy) density is defined as follows:
\[
 {\cal H}=\pi (x)\partial_0\Psi (x)+\left(\partial_0\overline{\Psi}
(x)\right)\overline{\pi}(x)-{\cal L}
\]
\vspace{-8mm}
\begin{equation}
 \label{24}
\end{equation}
\vspace{-8mm}
\[
=\frac{i}{2}\overline{\Psi}(x)\Gamma_4\partial_0\Psi (x)-\frac{i}{2}
\left(\partial_0\overline{\Psi}(x)\right)\Gamma_4\Psi (x) .
\]
It is easy to verify that ${\cal H}=T_{44}$, where
\[
T^c_{\mu\nu}= \frac{1}{2}\left(\partial_\nu\overline{\Psi}(x)\right)\Gamma_\mu
\Psi (x)-\frac{1}{2}\overline{\Psi} (x)\Gamma_\mu
\partial_\nu\Psi (x)
\]
is the canonical energy-momentum tensor. One may verify, with the
help of Eq.(10),(13), that this tensor is conserved:  $\partial_\mu
T^c_{\mu\nu}=0$.

\section{Quantization}

From the standard relation $[\Psi_M(\textbf{x},t),\pi_N
(\textbf{y},t)]= i \delta_{MN} \delta(\textbf{x}-\textbf{y})$, we
arrive, using Eq.(22), at simultaneous quantum commutators
\begin{equation}
\left[\left(\Psi(\textbf{x},t)\right)_M,
\left(\overline{\Psi}(\textbf{y},t)\Gamma_4 \right)_N\right ]=
2\delta_{MN}\delta(\textbf{x}-\textbf{y}) .\label{25}
\end{equation}
From Eq.(25),(7),(11), one obtains non-zero field commutators:
\begin{equation}
\left[B^\ast(\textbf{x},t),B_4(\textbf{y},t)\right ]=
2m\delta(\textbf{x}-\textbf{y}),~~\left[B_{[m4]}(\textbf{x},t),B^\ast_n(\textbf{y},t)
\right]=2 m\delta_{mn}\delta(\textbf{x}-\textbf{y}).\label{26}
\end{equation}

Let us consider solutions to Eq.(10) with definite spin, spin
projections, energy and momentum for two mass states in the form of
plane waves:
\begin{equation}
\Psi_{s,\tau}^{(\pm)}(x)=\sqrt{\frac{m_\tau^2}{p_0 V
M}}U_{s,\tau}(\pm p)\exp(\pm ipx) , \label{27}
\end{equation}
where $V$ is the normalization volume. We use the normalization
conditions
\begin{equation}
\int_V \overline{\Psi}^{(\pm)}_{s,\tau}(x)\Gamma_4
\Psi^{(\pm)}_{s',\tau }(x)d^3 x=\pm \delta_{ss'} ,~~~~\int_V
\overline{\Psi}^{(\pm)}_{s,\tau}(x)\Gamma_4 \Psi^{(\mp)}_{s',\tau
}(x)d^3 x=0 , \label{28}
\end{equation}
where
$\overline{\Psi}^{(\pm)}_{s,\tau}(x)=\left(\Psi^{(\pm)}_{s,\tau
}(x)\right)^+ \eta$. With the help of normalization conditions (28),
one obtains relations for the functions $U_{s,\tau}(\pm p)$:
\[
\overline{U}_{s,\tau}(\pm p)\Gamma_\mu U_{s',\tau}(\pm
p)=\mp\frac{iMp_\mu}{m_\tau^2}\delta_{ss'}
,~~~~\overline{U}_{s,\tau}(\pm p) U_{s',\tau}(\pm p)=\delta_{ss'},
\]
\vspace{-8mm}
\begin{equation}
\label{29}
\end{equation}
\vspace{-8mm}
\[
\overline{U}_{s,\tau}(\pm p)\Gamma_4 U_{s',\tau}(\mp p)=0.
\]

In the second quantized theory the field operator may be written as
\begin{equation}
\Psi_\tau(x)=\sum_{p,s}\left[a_{s,p}\Psi^{(+)}_{s,\tau}(x) +
b^+_{s,p}\Psi^{(-)}_{s,\tau}(x)\right] , \label{30}
\end{equation}
where positive and negative parts of the wave function are defined
by Eq.(27). The creation and annihilation operators of
particles $a^+_{s,p}$, $a_{s,p}$ and antiparticles $b^+_{s,p}$, $b_{s,p}$
satisfy the commutation relations:
\[
[a_{s,p},a^+_{s',p'}]=\delta_{ss'} \delta_{pp'}
,~~~[b_{s,p},b^+_{s',p'}]=\delta_{ss'} \delta_{pp'},
\]
\vspace{-8mm}
\begin{equation}
\label{31}
\end{equation}
\vspace{-8mm}
\[
[a_{s,p},a_{s',p'}]=[b_{s,p},b_{s',p'}]=[a_{s,p},b_{s',p'}]=[a_{s,p},b^+_{s',p'}]=0 .
\]
With the help of Eq.(27),(28),(30),(31), one obtains the Hamiltonian
\begin{equation}
H=\int_V {\cal H}d^3 x=\sum_{s,p}p_0\left(a^+_{s,p} a_{s,p}+b_{s,p}
b^+_{s,p}\right) . \label{32}
\end{equation}
We find from Eq.(30),(31) commutation relations as follows:
\begin{equation}
[\Psi_{\tau M(x)},\Psi_{\tau N}(x')]=[\overline{\Psi}_{\tau M}(x),
\overline{\Psi}_{\tau N}(x')] =0, \label{33}
\end{equation}
\begin{equation}
[\Psi_{\tau M}(x),\overline{\Psi}_{\tau N}(x')]=N_{\tau MN}(x,x'),
\label{34}
\end{equation}
\[
N_{\tau MN}(x,x')=N^+_{\tau MN}(x,x')-N^-_{\tau MN}(x,x') ,
\]
\begin{equation}
N^+_{\tau MN}(x,x')=\sum_{s,p}\left(\Psi^{(+)}_{s,\tau}(x)\right)_M
\left(\overline{\Psi^{(+)}_{s,\tau}}(x')\right)_N  ,\label{35}
\end{equation}
\[
N^-_{\tau MN}(x,x')=\sum_{s,p}\left(\Psi^{(-)}_{s,\tau}(x)\right)_M
\left(\overline{\Psi^{(-)}_{s,\tau,}}(x')\right)_N .
\]
From Eq.(30), one arrives at
\begin{equation}
N^\pm_{\tau MN}(x,x')=\sum_{s,p}\frac{m_\tau^2}{p_0
VM}\left(U_{s,\tau}(\pm p)\right)_M\left(\overline{U}_{s,\tau}(\pm
p)\right)_N\exp [\pm ip(x-x')] ,
 \label{36}
\end{equation}
Taking into account Eq.(17),(20), and the relation $p^2=-m_\tau^2$,
from Eq.(35), we obtain:
\[
N^\pm_{\tau MN}(x,x')=\sum_{p}\left\{
\frac{i(m+m_0)^4m_\tau^2\widehat{p}\left(i\widehat{p}\mp
M\right)}{2p_0VMm^2m_0^2 \left(m_\tau^4-m^2m_0^2\right)}\left[
\widehat{p}^2+
 \frac{m_\tau^4}{(m+m_0)^2}\right]\right\}_{MN}
\]
\begin{equation}
\times \exp [\pm ip(x-x')]= \biggl\{
\frac{(m+m_0)^4m_\tau^2\left(\pm\Gamma_\mu\partial_\mu\right)
\left(\pm\Gamma_\mu\partial_\mu  \mp M\right)}{Mm^2m_0^2
\left(m_\tau^4-m^2m_0^2\right)}\label{37}
\end{equation}
\[
\times \left[
 \frac{m_\tau^4}{(m+m_0)^2}- \left(\Gamma_\mu\partial_\mu\right)^2\right]\biggr \}_{MN}
 \sum_{p}\frac{1}{2p_0V}\exp[\pm ip(x-x')] .
\]
Using the singular functions \cite{Ahieser}
\[
\Delta_+(x)=\sum_{p}\frac{1}{2p_0V}\exp
(ipx),~~~~\Delta_-(x)=\sum_{p}\frac{1}{2p_0V}\exp (-ipx),
\]
\[
\Delta_0 (x)=i\left(\Delta_+(x)-\Delta_-(x)\right),
\]
one finds from Eq.(35),(37) the equation
\[
N_{\tau MN}(x,x')=-i\biggl\{
\frac{(m+m_0)^4m_\tau^2\left(\Gamma_\mu\partial_\mu\right)
\left(\Gamma_\mu\partial_\mu- M\right)}{Mm^2m_0^2
\left(m_\tau^4-m^2m_0^2\right)}
\]
\vspace{-6mm}
\begin{equation} \label{38}
\end{equation}
\vspace{-6mm}
\[
\times\left[
 \frac{m_\tau^4}{(m+m_0)^2}- \left(\Gamma_\mu\partial_\mu\right)^2\right]\biggr
 \}_{MN}\Delta_0 (x-x') .
\]
If the points $x$ and $x'$ are separated by the space-like interval
$(x-x')>0$, the commutator $[\Psi_M (x),\overline{\Psi}_N(x')]$
vanishes due to the properties of the function $\Delta_0 (x)$
\cite{Ahieser}. With the help of Eq.(15), one may verify that the
equation
\begin{equation}
\left(\Gamma_\mu\partial_\mu +M\right)N^\pm_{\tau}(x,x')=0 .
 \label{39}
\end{equation}
is valid.  The vacuum expectation of chronological pairing of
operators (the propagator) is defined as follows:
\[
\langle T\Psi_{\tau M}(x)\overline{\Psi}_{\tau
N}(y)\rangle_0=N^c_{\tau MN}(x-y)
\]
\vspace{-6mm}
\begin{equation} \label{40}
\end{equation}
\vspace{-6mm}
\[
=\theta\left(x_0 -y_0\right)N^+_{\tau MN}(x-y)+\theta\left(y_0
-x_0\right)N^-_{\tau MN}(x-y) ,
\]
where $\theta(x)$ is the theta-function. From Eq.(37), we obtain
\[
\langle T\Psi_{\tau M}(x)\overline{\Psi}_{\tau N}(y)\rangle_0
=\biggl\{ \frac{(m+m_0)^4m_\tau^2\left(\Gamma_\mu\partial_\mu\right)
\left(\Gamma_\mu\partial_\mu- M\right)}{Mm^2m_0^2
\left(m_\tau^4-m^2m_0^2\right)}
\]
\vspace{-6mm}
\begin{equation} \label{41}
\end{equation}
\vspace{-6mm}
\[
\times\left[
 \frac{m_\tau^4}{(m+m_0)^2}- \left(\Gamma_\mu\partial_\mu\right)^2\right]\biggr
 \}_{MN}\Delta_c (x-y).
\]
The function $\Delta_c (x-y)$ is given by
\begin{equation}
\Delta_c (x-y)=\theta\left(x_0
-y_0\right)\Delta_+(x-y)+\theta\left(y_0 -x_0\right)\Delta_-(x-y) .
\label{42}
\end{equation}
It follows from Eq.(41) that the propagator with $\tau=0$ possesses
the opposite sign compared with the case $\tau=1$. This confirms
that the state of the field with spin-0 is the ghost.

\section{Conclusion}

We have considered the fields possessing two masses,
$m$, $m_0$ with spin one and zero, i.e multi-spin 1,0. The
propagators of fields in the second order and first order
formulations are obtained. We found the density matrix and performed
canonical quantization of bosonic fields in the first order
formalism. The commutation relations and the vacuum expectation of
chronological pairing of operators obtained allow us to investigate
quantum processes with bosonic fields. It should be
stressed also that ghosts (k-essence, Phantom ) are widely used in
modern cosmology \cite{Mukhanov}. Therefore, the presence of spin-0
state of the fields may be used in some cosmological models.

\end{document}